\documentclass[]{article}
\usepackage{geometry}
\usepackage{lineno,hyperref}
\usepackage{float}
\usepackage{bm}
\usepackage{tabularx}
\usepackage{graphicx}
\usepackage[utf8]{inputenc}
\usepackage{epstopdf}
\usepackage[caption = false]{subfig}
\usepackage{amsmath}
\usepackage{cleveref}
\usepackage{siunitx}
\usepackage{multicol, caption}
\modulolinenumbers[5]
\usepackage{lipsum}
\newenvironment{Figure}
  {\par\medskip\noindent\minipage{\linewidth}}
  {\endminipage\par\medskip}
\usepackage{filecontents}
\usepackage[numbers]{natbib}

\begin{document}

\title{The thin lens equation in elasticity: imaging with gradient index phononic crystals}



\author{
P. H. Beoletto$^1$, F. Nistri$^1$, A.S. Gliozzi$^1$, N.M. Pugno$^{2, 3}$, F. Bosia$^1$\vspace{0.5cm}\\ 
\small{$^1$ Department of Applied Science and Technology, Politecnico di Torino,}\\ \small{C.so Duca degli Abruzzi, 24 - 10129, Torino, Italy.}\\
\small{$^2$ Laboratory for Bio-inspired, Bionic, Nano, Meta Materials and Mechanics,  Department of Civil,}\\
\small{ Environmental and Mechanical Engineering, University of Trento, 38123 Trento, Italy.}\\
\small{$^3$ School of Engineering and Materials Science, Queen Mary University of London, Mile End Road,}\\
\small{London E1 4NS, United Kingdom.}
}
\date{}

\maketitle
\begin{abstract}
Many works in elasticity have exploited the concept of gradient index (GRIN) lenses, borrowed from optics, for wave focusing and control. These effects are particularly attractive for cloaking, absorption or energy harvesting applications. Despite their potential, current lens designs suffer from limitations, mainly related to the difficulty in imaging point-like sources. Here, we exploit an alternative GRIN lens design, which enables a one-to-one correspondence between input and output phase, and allows to determine the focal length using the well-known thin lens equation, effectively establishing the elastic equivalent of the convex lens in optics. This is demonstrated analytically, obtaining a bijective relation between the location of a point-like source and its image, and the results are confirmed numerically and experimentally in an aluminium plate, where the lens is realized by introducing rows of circular cavities of variable diameters. Moreover, a proof-of-concept experiment demonstrates the possibility to image sources of flexural waves at the centimetre scale with subwavelength resolution. This research can extend applications of elastic GRIN lenses to new fields such as imaging and non-destructive testing, where the location of defects can be identified by focusing the scattered field. Multiple sources can be imaged simultaneously, and the combined effect of multiple lenses can also be used to design more complex systems, opening new possibilities in the technological exploitation of elastic wave manipulation.

\end{abstract}

\clearpage
\begin{multicols}{2}
\section*{Introduction}
The mathematical analogy between the laws governing electromagnetism and linear elasticity has often led  researchers to draw inspiration from photonic devices to obtain similar effects in the field of elastic wave propagation. The first full band-structure calculations for periodic elastic composites were based on studies on photonic crystals \cite{kushwaha1993acoustic} and, from then on, many aspects of interest for phononic crystals have followed a similar  trend\cite{pennec2010two}. The quest for wide frequency band-gaps exploiting Bragg or local resonances\cite{dal2021band}\cite{chen2016low}\cite{krushynska2017coupling}\cite{armenise2010phononic}\cite{park2024design}, the design of topological waveguides \cite{jiao2022observation}\cite{miniaci2021design}\cite{yu2018elastic}\cite{mousavi2015topologically}, and gradient-index acoustics \cite{lin2009gradient} highlight several similarities in the theory of elastic metamaterials with their optical counterpart. While Gradient Index lenses can also be designed by modulating the thickness of the medium supporting the propagation \cite{lee2021singular}\cite{zareei2018continuous}\cite{fuentes2021design}\cite{climente2014gradient}, Gradient Index Phononic Crystals (GRIN PCs) exploit local variations of the unit cell geometry to design a refractive index profile that induces the desired effect on the propagating acoustic wave in a specific frequency band: in subwavelength conditions (i.e., low frequencies), these devices behave as homogeneous non-dispersive materials whose effective properties are tuned by locally changing the geometry of the units \cite{jin2019gradient}. 
While wave focusing can be achieved by exploiting negative refraction at frequencies above the first complete band gap \cite{yang2004focusing}\cite{imamura2004negative}\cite{zhu2014negative}\cite{sukhovich2009experimental}\cite{christensen2012anisotropic}, changing the radius of the holes modifies the phase velocity of all the modes, allowing to work in wide bands \cite{lin2009gradient} at low frequencies.

In the past decade, much attention has been dedicated to the idea of exploiting GRIN lenses to focus elastic waves by acting on the local phase of the propagating wave to obtain the desired interference effects\cite{cui2019multi}\cite{antonacci2022planar}. A typical refractive index profile chosen for this kind of device is in the form of a hyperbolic secant\cite{wu2011focusing}, inspired from previous studies in gradient index optics \cite{akmansoy2008graded}\cite{huang2010tunable}\cite{gomez2012gradient}; this profile allows to focus an incoming plane wave in a focal point and has mainly been used for energy harvesting applications \cite{hyun2019gradient}\cite{zhao2021broadband}\cite{tol2016gradient}\cite{tol20193d}\cite{zega2022planar}\cite{lee2023achromatic}. Several works propose the use of circular Lunenburg lenses to obtain similar effects of plane wave focusing \cite{tol2017phononic}\cite{jin2016gradient}\cite{park2020recent}\cite{ma2022energy}. Both options allow unprecedented subwavelength focusing, unattainable with traditional materials in elasticity. 

Despite these attractive features, GRIN lenses are limited in their applicability. Hyperbolic secant profiles \cite{cui2019multi}\cite{antonacci2022planar} are monodirectional and they are only designed to focus incoming plane waves. Lunenburg lenses can be seen as omnidirectional\cite{tol2017phononic}, but their focusing properties are also limited to plane waves, and they do not display the ability to image finite objects. Maxwell's fish-eye elastic lenses have also been proposed\cite{lefebvre2015experiments}\cite{lefebvre2023subwavelength}, drawing inspiration from optics\cite{merlin2011maxwell}\cite{leonhardt2009perfect}\cite{ma2012subwavelength}: their dynamics is able to refocus the field generated by point-like sources, but this effect is limited to sources located at the boundaries of the crystal.

In this paper, we propose an alternative GRIN lens design for out-of-plane flexural waves, based on the concept of ``partitions" introduced in \cite{hyun2020partitioned}: by separating neighbouring rows of the PC, we prevent the waves propagating in one layer from interacting with those in adjacent ones, allowing to establish a one-to-one correspondence of the output phase with the input one. Thus, introducing a parabolic refractive index profile in the direction transverse to the propagation, we establish the desired focal length.
The fact that the phase delay induced by the lens is independent of the profile of the incoming wave allows us to develop an analytical model in paraxial approximation showing that this device mimics the simplest and most widely used optical device, \textit{i.e.} the convex lens, and conforms to Snell's law for thin lenses, also known as the ``Lens maker's equation" \cite{halliday2013fundamentals}, for which imaging conditions are obtained at a specific distance from the lens for a given source location. Given the similarities with the optical convex lens\cite{ma2022acoustic}, the device proposed in this work is referred to as ``positive flat lens", as it maintains the same converging and imaging properties without requiring a curved shape.

The device presented in this work can go beyond the limitations of existing GRIN lenses, allowing unprecedented flexibility in terms of applications. Exploiting the possibility of establishing a bijective relation between the object location and its image, it can map the pre-lens plane to the post-lens one. Thus, the novelty of this work lies in the demonstration of effective imaging of point-like sources, regardless of their shape and location.
Our approach includes a theoretical analytical to determine the focal properties of the lens, a numerical verification of the full-field elastic wave propagation problem and an experimental realization and characterization of the lens in lab tests. 


\section*{Results}

\subsection*{Design of positive flat lenses}
\begin{figure*}[t]
    \centering
    \includegraphics[width=\textwidth]{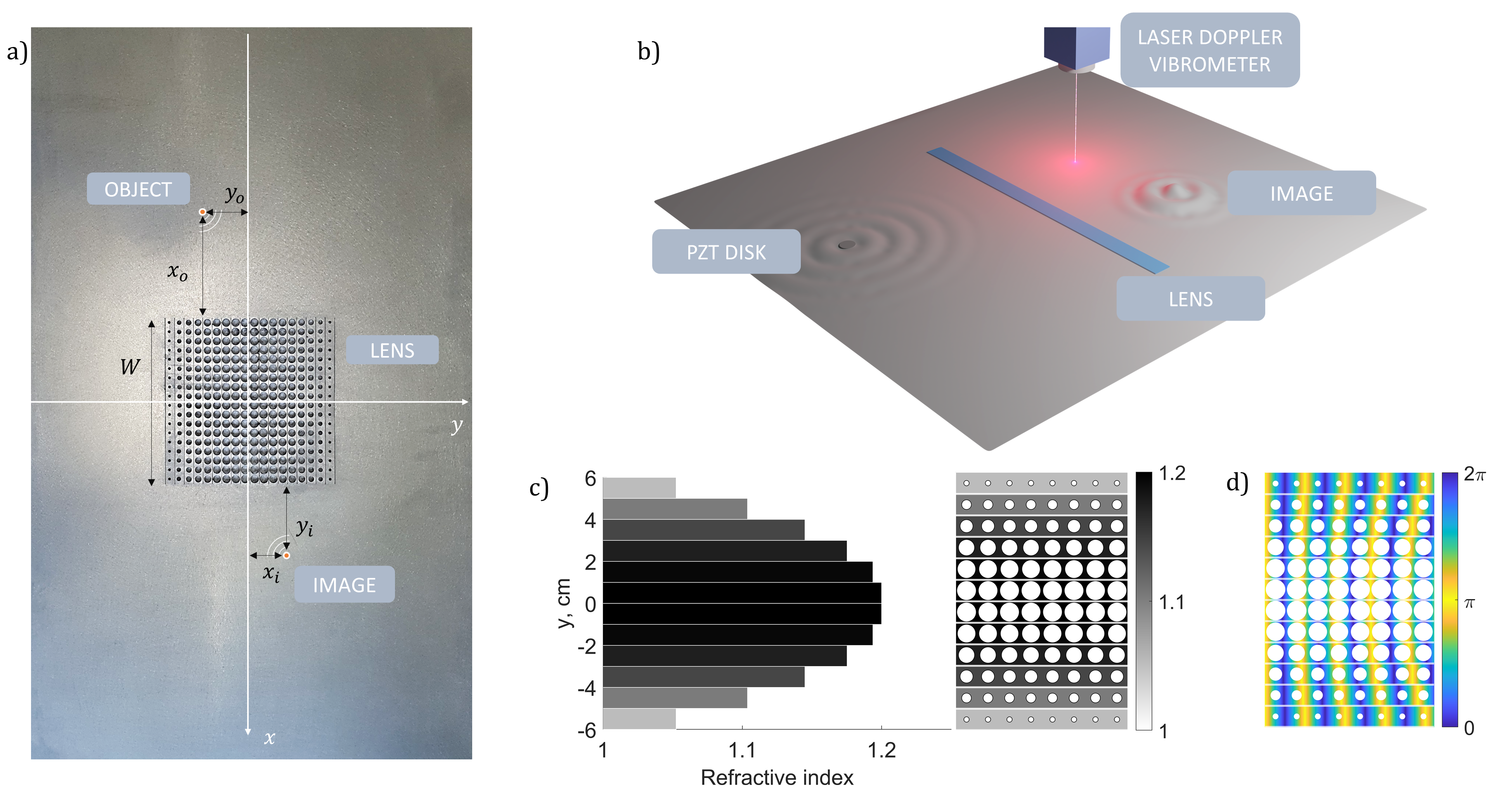}
    \caption{\textbf{a)} GRIN lens engraved in an aluminium plate that was used for the experiments. \textbf{b)} Render of the experimental set up: a piezoelectric disk is used to generate a propagating flexural wave and a Laser Doppler Vibrometer is used to scan over the post-lens area and detect the time dependent out-of-plane displacement point by point. \textbf{c-d)} Representation of the effect of the grading of the refractive index on the phase of the flexural waves propagating inside the lens: larger radii of the enclosures correspond to higher refractive indexes and shorter wavelengths.}
    \label{fig:1}
\end{figure*}

The lens design is based on a partitioned phononic crystal \cite{hyun2020partitioned}, ensuring wave propagation in each ``channel" to be independent of the others. Thus, the phase delay imposed at a specific position $y$ of the lens depends only on the refractive index $n(y)$ of that specific line, based on the bijective relation $\Delta\phi(y) = W k_0 n(y)$, where $W$ is the length of the acoustic path inside the lens and $k_0$ is the wavevector in the original material. A parabolic refractive index profile leads to a parabolic post-lens phase profile. In this work, we consider a lens (fig. \ref{fig:1}.a-b) with focal length $f=\SI{15}{\centi\meter}$ and working frequency $\nu = \SI{20}{\kilo\hertz}$. The supporting medium is a thin aluminium plate of area $\qtyproduct[product-units = power]{100x60}{\centi\meter}$ and thickness $d = \SI{3}{\milli\meter}$, the partitions are $\SI{0.2}{\milli\meter}$-wide slits separating the layers, and the unit cells of the phononic crystal are squares in the $xy$ plane with side length $a = \SI{1}{\centi\meter}$ and circular holes (fig. \ref{fig:2}.a), whose radius $r$ determines the refractive index (fig. \ref{fig:1}.c-d). 

Eigenfrequency simulations imposing 2D Bloch-Floquet quasi-periodicity in the $xy$ plane are conducted for 10 values of the filling factor $F = \frac{2r}{a}$, sampling the Brillouin zone in the $\Gamma-X$ direction.  Full 3D FEM simulations are carried out via the Finite Element solver COMSOL Multiphysics. Aluminium is modeled with the following parameters in the linear elastic constitutive law: density $\rho=\SI{2700}{\kilo\gram\per\cubic\meter}$, Young's modulus $E = \SI{70}{\giga\pascal}$ and Poisson's ratio $\nu = 0.33$. The cell is meshed by means of tetrahedral elements with a maximum size of $\SI{0.55}{\milli\meter}$, which are automatically created by the software in order to preserve the geometry of the circular enclosure. The dispersion relation of the first anti-symmetric bending mode $A_0$ undergoes a red-shift with its effect intensifying  with an increasing filling factor (fig. \ref{fig:2}.b). Assuming that the frequency $\nu$ of this flexural mode has a quadratic dependence on the wavevector \cite{love1888xvi} $\nu = C(F)\cdot k^2$, where $C$ is a coefficient depending only on $F$, we obtain an expression for the refractive index $n(F) = k(F) / k_0 = \sqrt{C_0 / C(F)}$, i.e., $n$ increases monotonically with the filling factor (fig. \ref{fig:2}.c). This confirms that the relation between filling factor and phase delay is bijective. The refractive index is frequency independent for the $A_0$ mode, as the ratio between the wavevector in the crystal and in the supporting medium is unchanged for a given value of $C(F)$.
A lens with focal length $f$ can focus a plane wave at the coordinates $(f,0)$. The lens compensates the post-lens phase delay so that all acoustic paths contribute to constructive interference in this point. The phase delay imposed by the lens is $\Delta\phi(y) = \Phi_0 - 2\pi k_0 \sqrt{y^2+f^2}$, where $\Phi_0$ is a constant. This profile exhibits a dependence on the direction $y$ perpendicular to propagation that is the same as that of optical convex lenses, where the phase delay is imposed by the radius of curvature. In the case of the considered flat lens with fixed width $W$, the same effect is obtained with the refractive index profile  $n(y) = n_{max} - \sqrt{y^2+ f^2}/W$.

Considering the generic object location $(x_o,\ y_o)$ and image location $(x_i,\ y_i)$, the phase associated  with the acoustic paths of an elastic wave propagating from one point to the other is 
\begin{equation}
    \begin{aligned}
        \Phi_{tot}= &  2\pi k_0 \sqrt{y^2+x_o^2} + \Phi_0 \\
                    & - 2\pi k_0 \sqrt{y^2+f^2} +2\pi k_0 \sqrt{y^2+x_i^2}
    \end{aligned}
\end{equation}

Following a similar approach to the one used in optics \cite{born2013principles}, we can simplify the equation by applying a paraxial approximation, that holds when $y<<x_o,f,x_i$. In these conditions, we can write:
\begin{equation}
    2\pi k_0 \dfrac{y^2}{2} \left( \dfrac{1}{x_o} - \frac{1}{f} + \frac{1}{x_i}\right) = \Phi_{tot}
\end{equation}

The total phase at the location $(d_i, 0)$ is independent on the acoustic path only when
\begin{equation}
    \dfrac{1}{x_o}  + \frac{1}{x_i} = \frac{1}{f}
\end{equation}

This imaging condition is totally equivalent to the thin lens formula and it contains a one-to-one relationship between the $x$ coordinate of the object and that of its image, given the focal length. In optics, this law can be obtained when the lens is thin, \textit{i.e.} when the propagation inside the lens is almost uniaxial: in this approach, the equivalent concept is the partitioning of the lines of cells with different refractive indexes.

A final step to generalize this model is to extend the approach to out-of-axis source positions $(x_o, y_o)$. 

\begin{Figure}
    \centering
    \includegraphics[width=\linewidth]{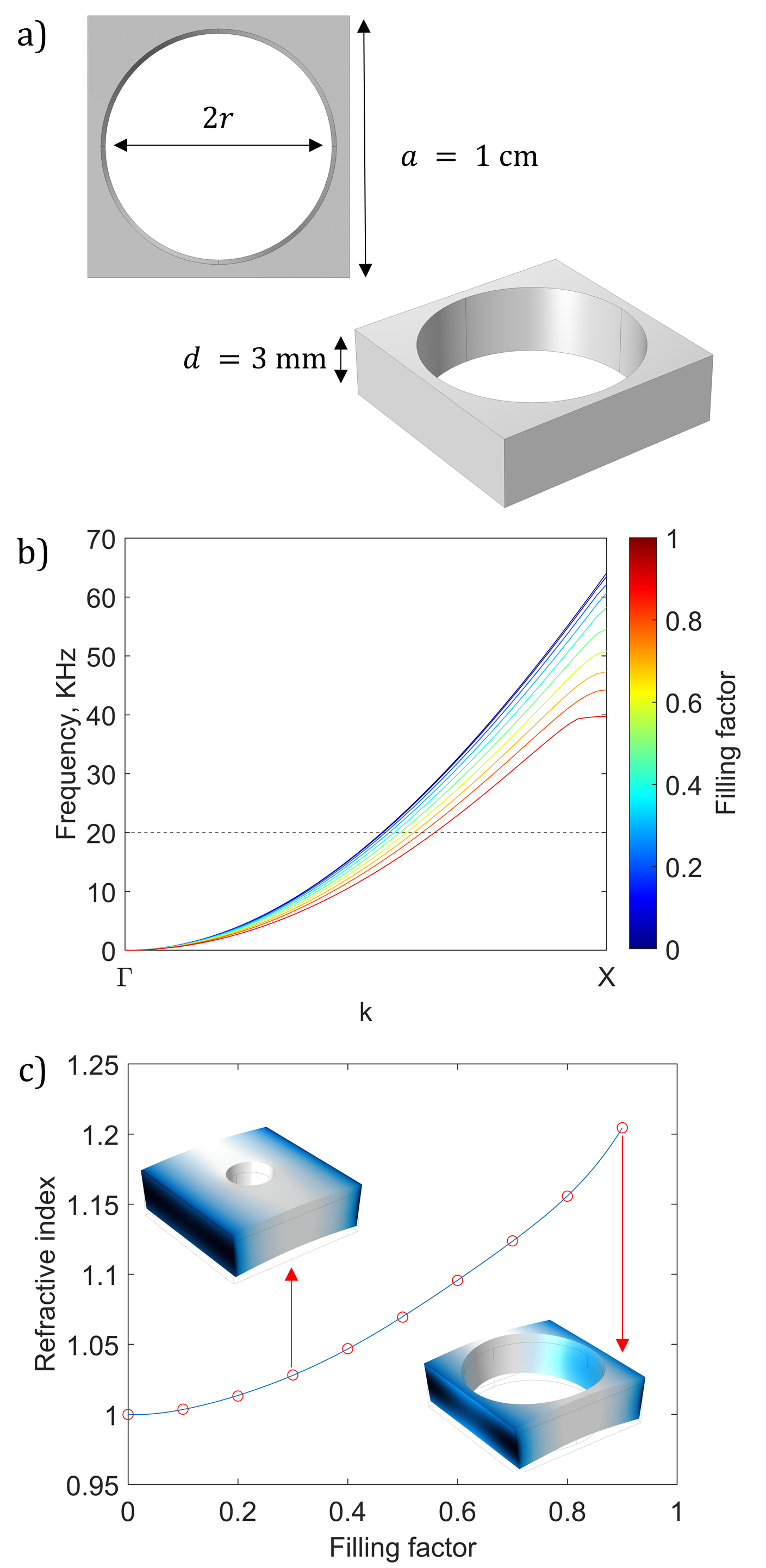}
    \captionof{figure}{ \textbf{a)} Representation of the unit cell: the filling factor is a parameter $\frac{2r}{a}$ that is used to quantify the grading. \textbf{b)} Red-shift of the $A_0$ mode with increasing filling factor: eigenfrequency simulations with Bloch-Floquet conditions were run on unit cells with filling factor ranging from $0.1$ to $0.9$. \textbf{c)} Bijective relation between the filling factor and the refractive index $n = \frac{k}{k_0}$.}
    \label{fig:2}
\end{Figure} 

The distance $r_{in}$ between the object and the lens at the coordinate $y$ is:
\begin{equation}
    r_{in} = \sqrt{(x_o)^2 + (y_o-y)^2} \approx d_o + \dfrac{(y-y_o)^2}{x_o}
\end{equation}

This leads to an input phase profile that can be written as:
\begin{equation}
    \phi_{in}(y) = 2\pi k_0 \left( x_o + \dfrac{1}{x_o} (y^2 + y_o^2 - 2 y y_o)\right)
\end{equation}

The same holds in the post-lens propagation phase term. To seek for the imaging conditions, we impose that the total phase associated to the propagation is a constant, $\Phi_{tot}$, independent of the acoustic path:
\begin{equation}
    \begin{aligned}
        \Phi_{tot}= &   2\pi k_0 \left(\dfrac{1}{d_o} - \frac{1}{f} + \frac{1}{d_i}\right) \dfrac{y^2}{2} \\
                    & +  2\pi k_0 \dfrac{1}{2}\left(\dfrac{y_o^2}{d_o}+\dfrac{y_i^2}{d_i} \right) -  2\pi k_0 \left(\dfrac{y_o}{d_o}+\dfrac{y_i}{d_i}\right)y
    \end{aligned}
\end{equation}

In the paraxial limit, that is described by this formula, it is possible to observe that the total phase is independent of the acoustic path (\textit{i.e.} $\Phi_{tot}$ is independent of $y$) when two imaging conditions are satisfied:

  \begin{equation}\label{eq:imaging1}
    \dfrac{1}{x_o}  + \frac{1}{x_i} = \frac{1}{f} 
    \end{equation}
    
  \begin{equation}\label{eq:imaging2}
    y_i = - \dfrac{x_i}{x_o}y_o
    \end{equation}

From these equations, we observe that positive flat lenses are capable of creating images of point-like objects: the $(x_o,\, y_o)$ space is mapped into the $(x_i,\, y_i)$ one, with a functional dependence whose only parameter is the focal length. Equation \ref{eq:imaging1} establishes the relationship between the position of the focal line and the distance of the source from the lens in the direction of propagation, while equation \ref{eq:imaging2} states that their off-axis positioning depends upon the ratio of these distances.

\subsection*{Determination of lens focal length}

The analytical model developed for the positive flat lens is initially verified through numerical and experimental approaches. The device is designed with a focal length of $\SI{15}{\centi\meter}$, which is the parameter to be validated before checking its imaging properties. To do this, numerical full-field frequency domain Finite Element (FE) 3D simulations are performed using Comsol Multiphysics.

\begin{Figure}
    \centering
    \includegraphics[width=\linewidth]{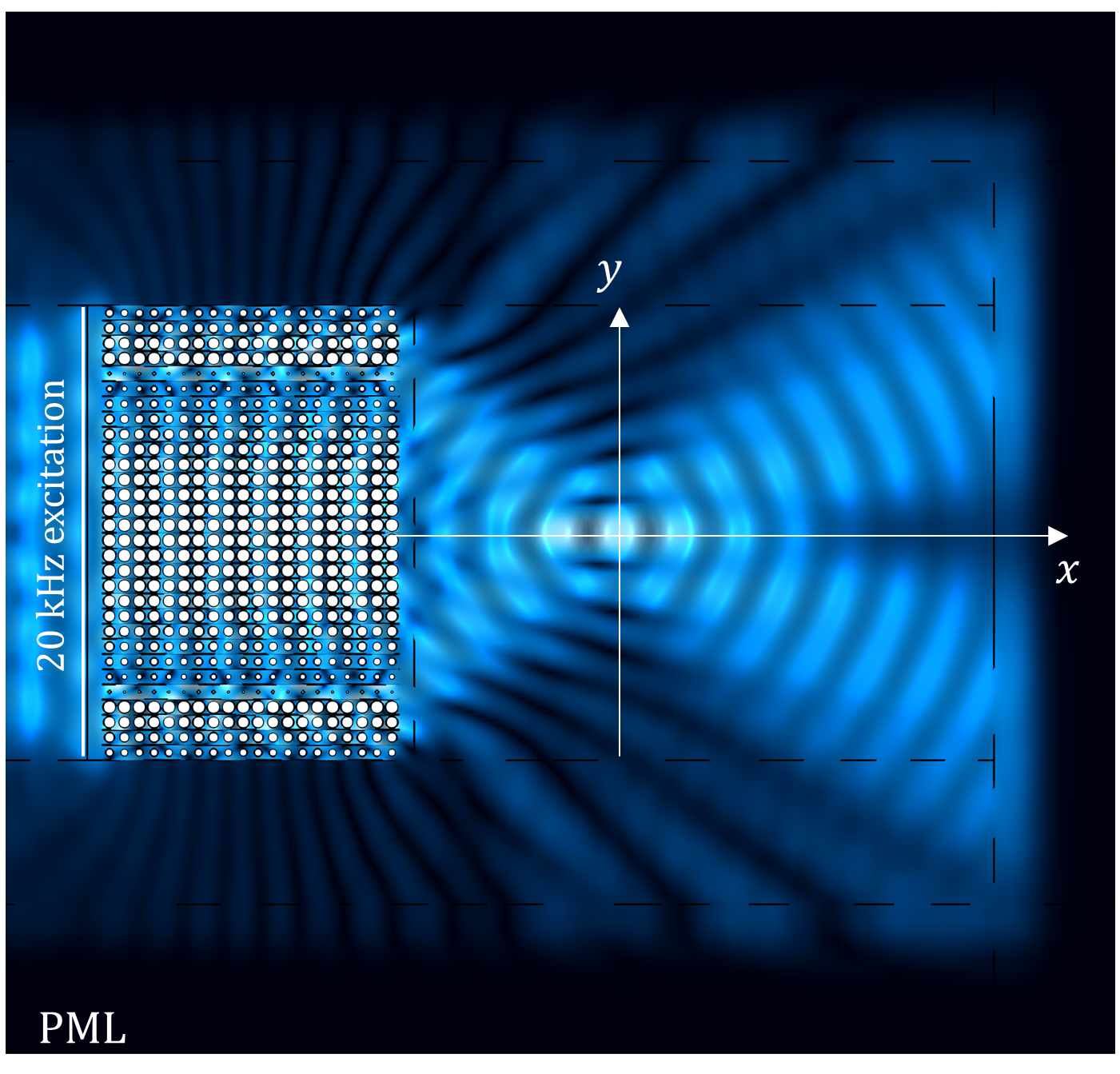}
    \captionof{figure}{Numerical simulation in frequency domain of the effect of a GRIN Lens designed to focus a $\SI{20}{\kilo\hertz}$ plane wave with a focal length $f =\SI{15}{\centi\meter}$ (low displacement amplitude in black, high displacement amplitude in white). }
    \label{fig:3}
\end{Figure}

\begin{Figure}
    \centering
    \includegraphics[width=\linewidth]{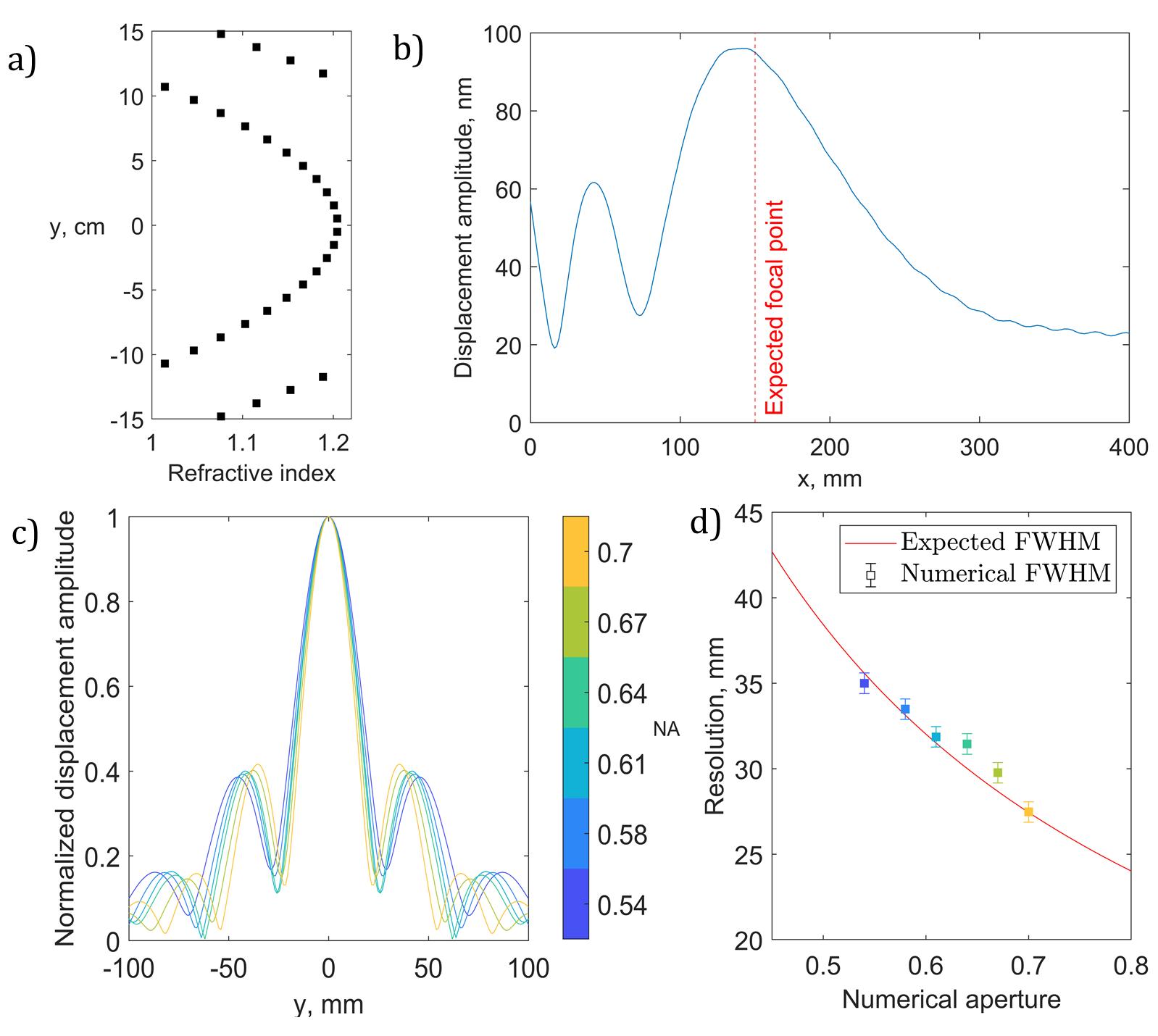}
    \captionof{figure}{\textbf{a)} Refractive index profile of the lens under study. \textbf{b)} Displacement amplitude profile in the propagation direction $x$ obtained from the numerical simulation in frequency domain. \textbf{c)} Normalized displacement distribution in the focal plane for various numerical aperture values of the lens. \textbf{d)} Resolution of the focal spot (calculated as its full width half maximum) compared to the expected resolution. }
    \label{fig:4}
\end{Figure} 
\begin{figure*}[hb]
    \centering
    \includegraphics[width=\textwidth]{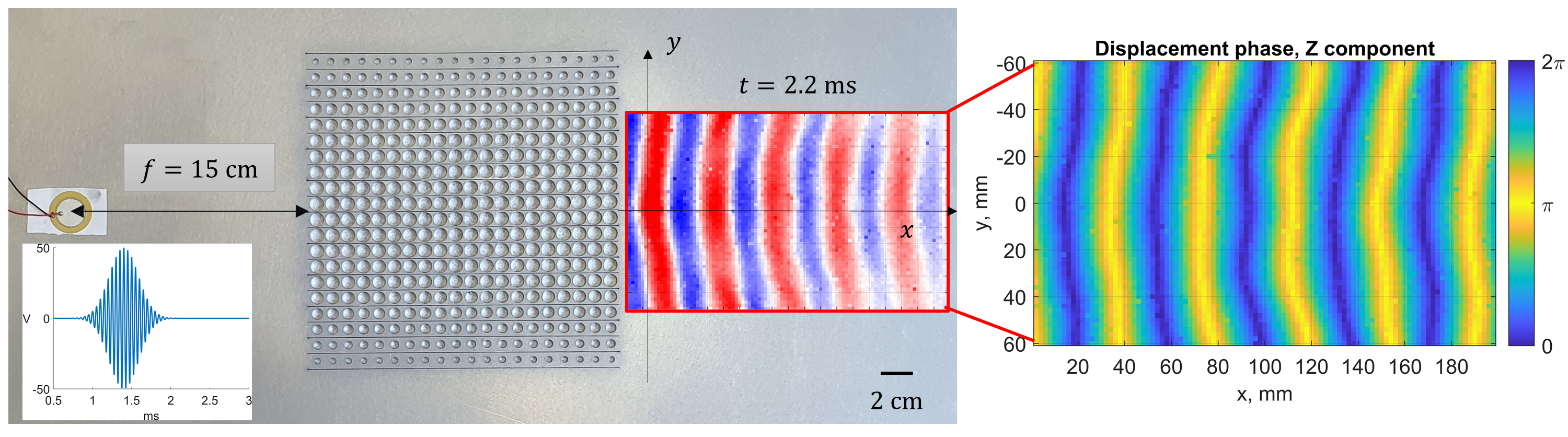}
    \caption{Experimental validation of the focalization properties of the lens: a piezoelectric source is placed in the focal point of the lens at a distance $f = \SI{15}{\centi\meter}$ and a snapshot of the plane wave propagating in the post-lens plane is shown. The input signal is a gaussian pulse centered at $\SI{20}{\kilo\hertz}$ with a $10\%$ bandwidth. The plot on the right shows the phase of the $\SI{20}{\kilo\hertz}$ component of the signal measured in the post-lens plane, with planar wavefronts.}
    \label{fig:5}
\end{figure*}  

Perfectly Matched Layer (PML) boundary conditions are implemented on the edges of the $\qtyproduct[product-units = power]{100x60}{\centi\meter}$ aluminium plate to minimize reflections at the boundaries: the PML is modeled with polynomial stretching,  featuring a scaling factor of $1$ and curvature parameter of $3$. 
The lens area is meshed with tetrahedral elements and the rest of the plate with 8-node hexahedral elements of maximum size $\SI{6}{\milli\meter}$. The lens is positioned at the center of the plate and a $\SI{20}{\kilo\hertz}$ excitation is imposed over a line to ensure a constant phase profile at the interface with the lens (\ref{fig:3}). The graded cell design gives rise to a the parabolic-like refractive index profile shown in \ref{fig:4}.a. 
The jump in refractive index is not perceived as a discontinuity since waves propagating in neighbouring channels do not interfere with each other and a phase delay equal to $2\pi$ at the lens end is equivalent to $0$. The first test is performed on a lens with an aperture of $\SI{30}{\centi\meter}$ and for this design the maximum displacement amplitude confirms the expected focal length value (\ref{fig:4}.b). The aperture of the lens has no influence on its focal length (see also Supplementary Material), but influences its resolution\cite{born2013principles}\cite{airy1835diffraction}: the Abbe diffraction limit defines the minimum resolvable size of an optical system as $d = \lambda / (2NA)$, where $NA = n \sin(\theta)$ represents the numerical aperture. To ascertain whether our lens operates as a diffraction-limited system, the same simulation is performed on the lens with lateral dimension ranging from $\SIrange{20}{30}{\centi\meter}$, \textit{i.e.} numerical aperture in the range $ 0.55 - 0.71$. The Full Width Half Maximum of the focal spot in each test (\ref{fig:4}.c-d) is compared with the expected value of the diffraction limit, demonstrating quantitative agreement with theoretical predictions.  Enhancing the system's resolution is possible by either increasing the aperture or using a higher refractive index in the post-lens plane.

\begin{figure*}[t]
    \centering
    \includegraphics[width=\textwidth]{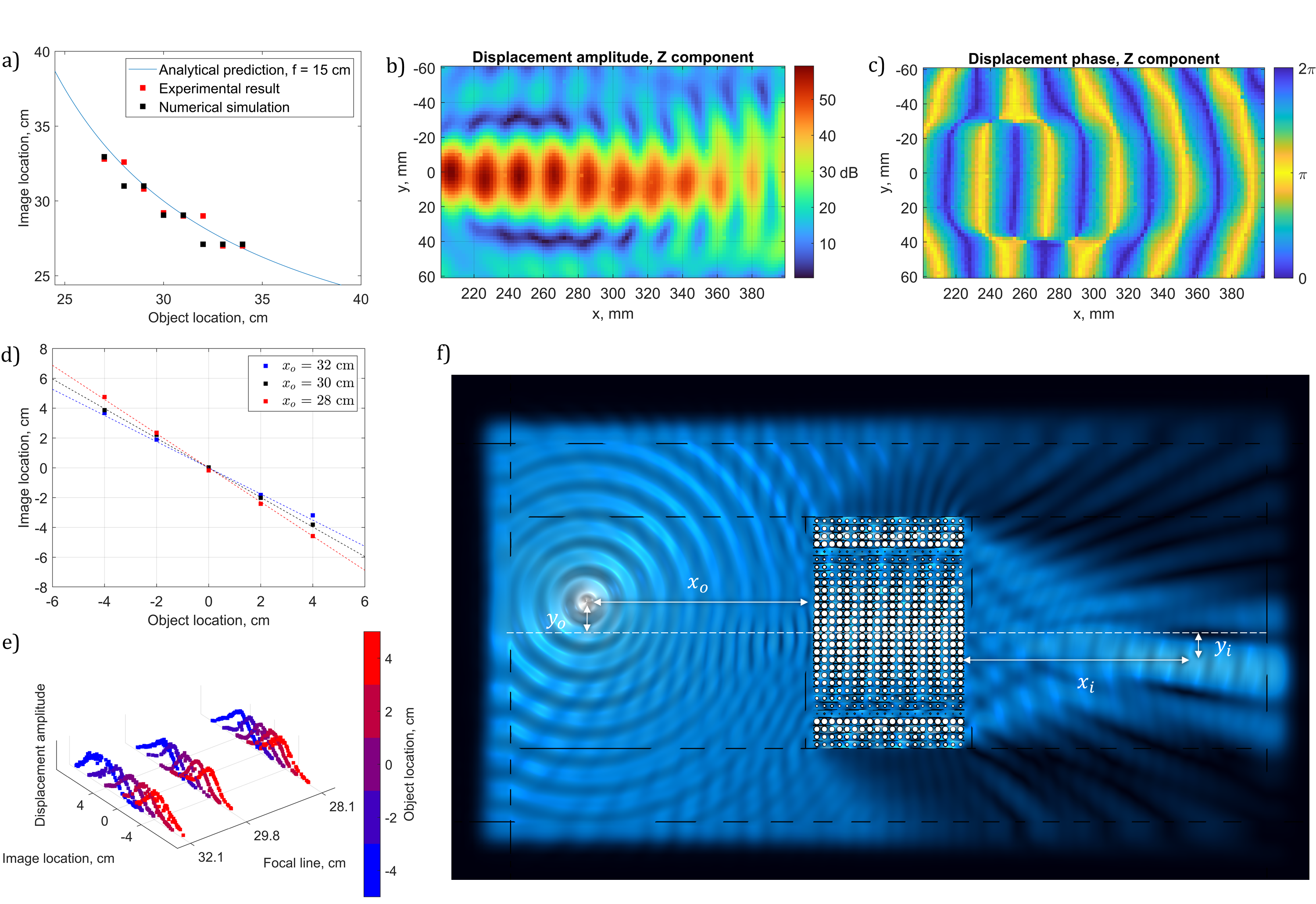}
    \caption{\textbf{a)} Comparison between analytical model, numerical simulations and experimental results for the imaging of a point-like source located at different distances from the lens: the image location corresponds to the point of maximum amplitude of the $\SI{20}{\kilo\hertz}$ component of the propagating wave along the central line in the post-lens plane. \textbf{b-c)} Post-lens map of the amplitude and phase of the $\SI{20}{\kilo\hertz}$ component of the signal generated with a source located in $x_o= \SI{37}{\centi\meter}$ and $y_o= \SI{0}{\centi\meter}$. \textbf{d-e)} Results of imaging experiments with the source located out of the central axis and the analytical expectations: three sets of experiments with $x_o = \SI{28}{\centi\meter}$, $\SI{30}{\centi\meter}$ and $\SI{32}{\centi\meter}$ are performed, measuring the output at the specific focal line $x_i$ and the peak location of each measurement is compared with the expected $y_i = -\frac{x_i}{x_o}y_o$. \textbf{f)} Numerical frequency domain simulation of the lens imaging a source located at $(x_o= \SI{30}{\centi\meter}, y_o= \SI{4}{\centi\meter})$.}
    \label{fig:6}
\end{figure*}

Additionally, experimental validation of the lens's focal length was undertaken: the lens used for the experiments is the $\SI{20}{\centi\meter}$ wide version. The lens design is fabricated on an alluminium thin plate via numerically-controlled laser cutting. Piezoelectric transducers are taped to the plate and used to excite elastic waves. The wavefields are measured using a Polytec scanning laser Doppler vibrometer (SLDV): the out-of-plane velocity of points belonging to a predefined grid is measured by moving the specimen in the $xy$ plane with two linear stages. Modeling clay is applied to the boundaries of the plate to partially reduce the back-scattering. The aim is to limit the steady-state effects on the wave-fields. Due to the finite dimensions of the supporting medium, generating a plane wave at the lens entrance, indicative of a source positioned infinitely distant, proved challenging; therefore, experiments are performed in a time-reversed configuration, by placing the object in the focal point and detecting the plane wave that is expected to be generated on the other side of the lens.  A piezoelectric transducer placed at a distance $f = \SI{15}{\centi\meter}$ from the lens is excited with a Gaussian pulse centered at $\SI{20}{\kilo\hertz}$ with $10\%$ bandwidth. This signal is sufficiently long to ensure spatial coherence\cite{pedrotti2017introduction} of the field at the lens entrance. A laser Doppler vibrometer is used to scan over the post-lens plane and record the time dependent out-of-plane displacement in each point (\ref{fig:5}), which confirmed the propagation of a plane wave. In order to check the radius of curvature of the wavefront, the signal in each point is Fourier transformed and the $\SI{20}{\kilo\hertz}$ component is analysed: apart from a small deviation at the edges, linked to the background signal propagating out of the lens, the phase profile indicates that the wavefronts are planar (\ref{fig:5} and Supplementary Material). Thus, both numerical simulations and experiments corroborate the analytical design's focal length prediction for the device.


\subsection*{Imaging point-like sources}

Positive GRIN lenses are not only capable of focalising energy: their working principle also enables imaging objects and reconstructing their location. Equations \ref{eq:imaging1} and \ref{eq:imaging2} show that the only parameter determining the relation between source and image position is the focal length $f$, whose value has been shown to be consistent with the design predictions both numerically and experimentally in the previous section. For this given value of the focal length, a one-to-one relation between the object and image location exists: in order to demonstrate the imaging capability of the device, numerical and experimental results are compared with analytical predictions.

From Eqs. \ref{eq:imaging1} and \ref{eq:imaging2}, it follows that $x_i = (x_of) / (x_o-f)$. This relation holds if the object is located at a distance larger than the focal length, since for smaller distances the field does not converge to an image. A set of values of $x_o$ around $2f$ is selected, ranging from $\SIrange{27}{34}{\centi\meter}$. Frequency domain FE simulations are carried out using Comsol Multiphysics, with a point excitation at a frequency of $\SI{20}{\kilo\hertz}$ for each value of $x_o$ and the same process is replicated experimentally using a piezoelectric disk as a source. The object is located along the central axis ($y_o=0$) and displacements in the post-lens area are studied on the same line. Numerically, the value of $x_i$ is determined as the distance from the lens of the maximum of the out-of-plane component of the displacement amplitude. In experiments, a Fourier transform is performed on the time domain signal measured in each point along the line and the amplitude of the $\SI{20}{\kilo\hertz}$ component of the signal is selected to determine the maximum value. The analytical predictions are confirmed both by the experiments and by the numerical simulations(fig. \ref{fig:6}.a). The numerical and experimental results indicate a step-like behavior in the relation between the object and image location: the values of $x_i$ appear to form a discrete set, attributable to the finite size of the supporting medium,  resulting in the presence of a back-scattered wave. The reflections from the sides of the plate create an interference pattern in steady-state conditions, whose maxima occur at intervals of $\frac{\lambda}{2}$ along the direction of propagation. This effect, superposed on the actual working principle of the device, is also accounted for in the simulations where the actual plate geometry is considered. Figure \ref{fig:6}.b shows the map of the experimentally detected displacement amplitude with the source located at $x_o= \SI{37}{\centi\meter}$: the fringes in this image correspond to the maxima of interference with the reflected wave. The map of the phase (fig. \ref{fig:6}.c) indicates that the image is located in the region where the converging wavefront begins to diverge.
The relation between off-axis positioning of the source and of the image is also verified. The lens has the capability to work both in magnification and demagnification. When the object is located at a distance $x_o<2f$ from the lens, the ratio $x_i/x_o>1$ applies: in this condition, distances are increased and the lens works in magnification. The opposite is true for $x_o>2f$, representing the demagnification condition, and for $x_o = 2f$ the object is precisely mirrored at the focal line ($y_i = - y_o$). Three sets of experiments are conducted both in a numerical and experimental environment, for $x_o = \SIlist{28; 30; 32}{\centi\meter}$, \textit{i.e.} $x_o<2f$, $x_o=2f$ and $x_o>2f$. In each condition, the amplitude profile is analyzed at the corresponding focal line with the source placed at the location $y_o$ ranging from $\SIrange[]{-4}{4}{\centi\meter}$. 
The position $y_i$ of the peak of the displacement amplitude of the $\SI{20}{\kilo\hertz}$ component determines the image location. Figure \ref{fig:6}.d demonstrates a significant correlation between the experimental results and the analytical predictions as outlined by equation \ref{eq:imaging2}.  As expected, distances are either magnified or demagnified depending on the operational state of the lens (fig. \ref{fig:6}.e). Numerical simulations (fig. \ref{fig:6}.f) confirm the experimental results (see also Supplementary Material). Therefore, the model for positive GRIN lenses is verified and the imaging properties of the device are fully respected for point-like sources.

\subsection*{Localization of objects}

The imaging capability of the positive GRIN lens can be extended from piezoelectric point-like sources to any object that interacts with the wavefield propagating in the pre-lens area. In this section, a localization technique for point-like defects is tested, based on the focalization and imaging properties of the lens.
\begin{Figure}
    \centering
    \includegraphics[width=\linewidth]{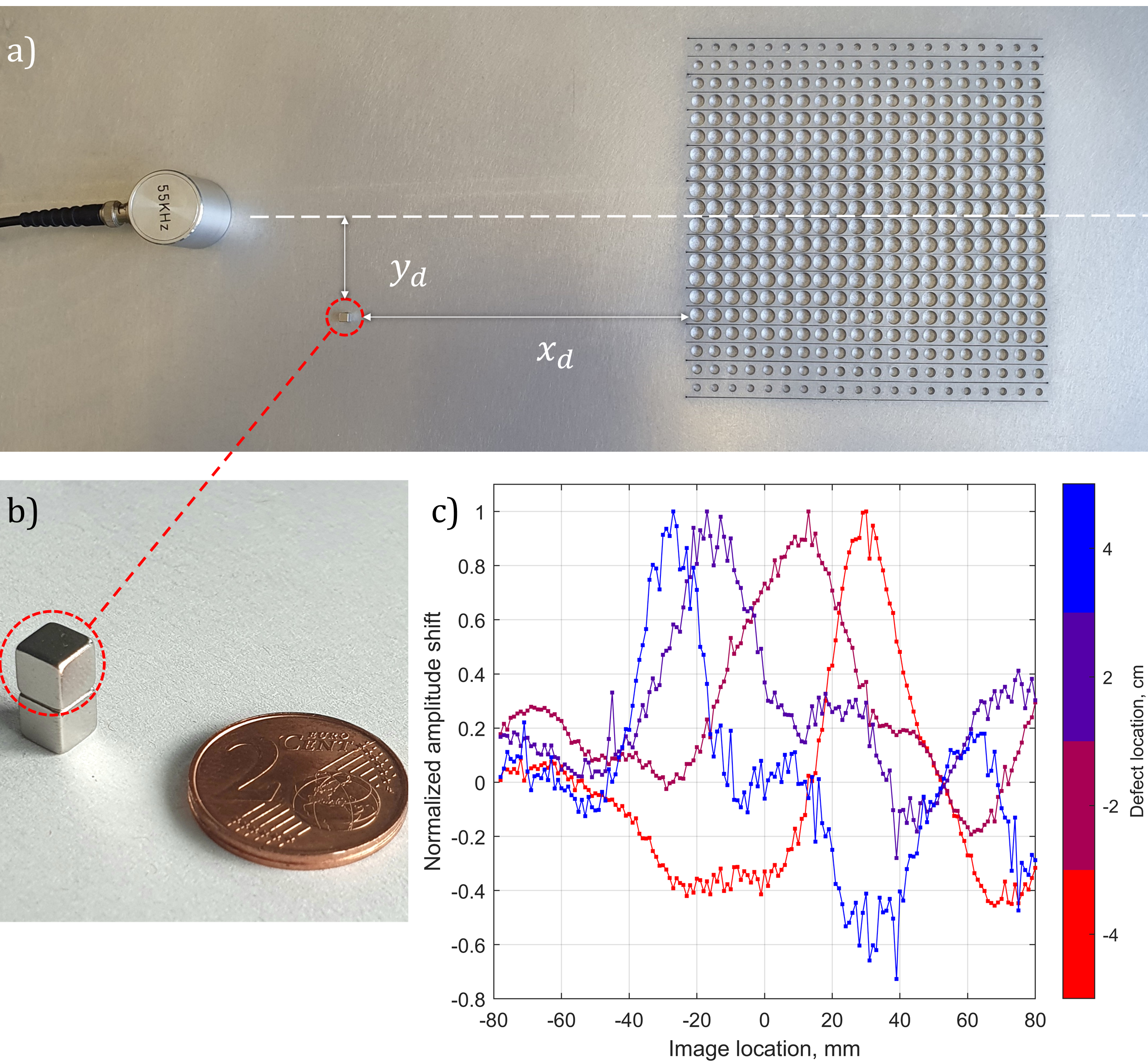}
    \captionof{figure}{\textbf{a)} Set up of the defect imaging experiment: the defect is located at $(x_d, y_d)$ and the signal in the focal line $x_i = \frac{f\cdot x_d}{x_d-f}$ is compared to a defect-less reference to check for variations in the amplitude profile. \textbf{b)} Magnetic cubes used as defect sources: the pressure on the plate generated by the magnetic attraction between the two cubes generates a local variation in the mechanical properties. \textbf{c)} Image of defects located in different positions observed in the focal line.}
    \label{fig:8}
\end{Figure} 

In the experiment, two cubic magnets (fig.\ref{fig:8}.a and \ref{fig:8}.b) are placed at a corresponding location on the upper and under sides of the plate, to create a local scattering obstacle\cite{wang2005scattering} to the wavefield at the location where they are placed. 
\begin{figure*}[t]
    \centering
    \includegraphics[width=\textwidth]{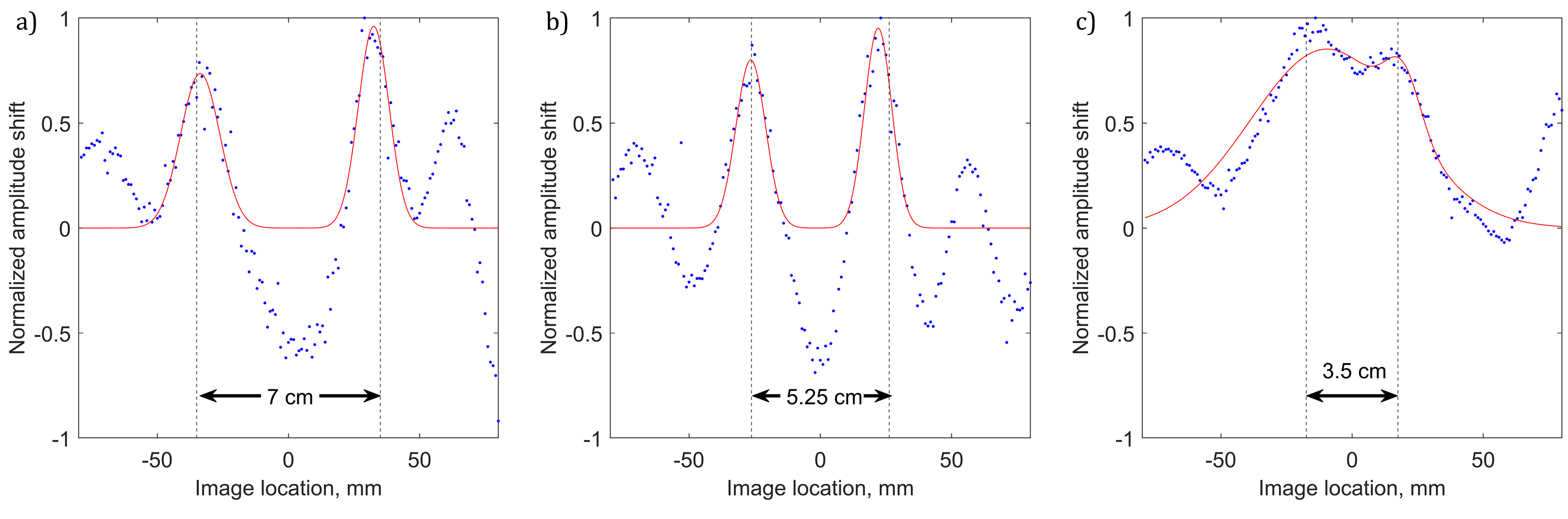}
    \caption{\textbf{a-b-c)} Peaks generated by the simultaneous presence of two defects located at a distance of $\SI{8}{\centi\meter}$, $\SI{6}{\centi\meter}$ and $\SI{4}{\centi\meter}$: in the last case the distance is comparable with the wavelength $\lambda= \SI{3.75}{\centi\meter}$ and the two defects can't be distinguished.}
    \label{fig:7}
\end{figure*}

A defect-less reference of the amplitude distribution of the signal is acquired,  enabling a comparison of amplitude variations when the defect is introduced. An initial set of experiments is run with the defect located at $x_d = \SI{32}{\centi\meter}$ and four different values of $y_d$ from $\SIrange{-4}{4}{\centi\meter}$ (fig.\ref{fig:8}.c).
 The amplitude shift is measured at the corresponding focal line $x_i = \SI{28.2}{\centi\meter}$, where the effect of the object is expected to be focalised. Figure \ref{fig:8}.c shows that the peak of the amplitude shift is localized in the position $y_i$, as predicted by the lens model. For the magnet placed at $y_d = \SI{-4}{\centi\meter}$, the expected image location is $y_i = \SI{3.5}{\centi\meter}$, as observed in the experiments.

A final experiment is performed to directly test the resolution of the system, by placing simultaneously two defects on the same line (fig.\ref{fig:7}.a-b-c).
The distance of the objects in the three experiments is respectively $\SIlist{8;6;4}{\centi\meter}$, with an expected distance at the image line of $\SIlist{7; 5.25; 3.5}{\centi\meter}$. The objects are well resolved in the first two configurations (fig.\ref{fig:7}), whilst they cannot be distinguished in the last one (fig.\ref{fig:7}.c). Several mechanisms need to be considered in evaluating the resolution of this system: first, the width of the lens, which defines the numerical aperture and  thus sets the diffraction limit. Second, the source used in this configuration generates a spherical wave: the field interacting with the object has non-zero components in the direction perpendicular to propagation as the magnets are placed off-axis. As the object approaches the central axis, the reduced value of $k_y$ in the scattered field increases the minimum distinguishable feature size, explaining why objects placed only $\SI{4}{\centi\meter}$ apart fail to be resolved.


\section*{Discussion}
In conclusion, we have proposed a design procedure for the elastic counterpart of the optical convex lens based on gradient index phononic crystals. 
The gradient index designs that are currently available in the domain of elasticity are effective in focusing energy but their working conditions are either limited to a specific direction of the incoming wave-field or location of the source. These limits are overcome by the device proposed in this work: the positive flat lens is the first phononic crystal based device that is able to create an image of a source of elastic waves regardless of its shape and location.
The analytical model for the positive flat lens predicts that the device can create images of objects in a location unequivocally determined by its focal length. Both numerical simulations and experiments performed with point-like sources confirm the design predictions in terms of both image location and resolution, ensuring full reliability to the physical model behind it. Moreover, the imaging capabilities of the lens have been tested in a scattering configuration: using mechanical defects generated by the local pressure of two magnets, the lens has been shown to be capable of simultaneously localising more than one object. This work opens a range of potential applications. The first is to exploit focusing for energy harvesting or for in material signal processing in low-power devices, since the simple design of the lens is amenable to miniaturization. Another possibility is related to the fact that the lenses can be used to magnify and demagnify objects, depending on their focal properties. This can be very useful, \textit{e.g.}, in non-destructive testing applications when the inspected area is very large or very small, or scarcely accessible. An appropriate lens design can allow to generate a scaled image of the measuring region in a convenient adjacent region. Finally, the combined effect of multiple lenses can also be used to design more complex systems for the control of flexural wave propagation, such as in elastic microscopes.

The convex lens, although simple in design, revolutionized the field of light manipulation and paved the way for advanced applications. The positive flat lens has the potential to also have a considerable impact and to open up new possibilities in the domain of elasticity.


\bibliographystyle{elsarticle-num}


\section*{Acknowledgements}
This project has received funding from the European Union’s Horizon 2020 FET Open (“Boheme”) under grant agreement No. 863179.

\section*{Author contributions}
P.H.B., A.S.G. and F.B. conceived the idea and coordinated the work; P.H.B. developed the analytical model and defined the design procedure; P.H.B. and A.S.G. performed the numerical simulations; A.S.G. built the experimental set up; P.H.B. and F.N. performed the experimental measurements and performed the data analysis; N.M.P supervised the work and provided funding. All the authors contributed to the manuscript writing.

\end{multicols}

\end{document}